\documentclass[12pt]{iopart}

\usepackage{graphicx}
\usepackage{fancyhdr}
\usepackage{amsfonts,amssymb,color}

\definecolor{red}{rgb}{1,0,0}

\begin{document}

\title[]{Anomaly free cosmological perturbations with generalised holonomy correction in loop quantum cosmology}

\author{Yu Han, Molin Liu}

\address{College of Physics and Electrical Engineering, Xinyang Normal University, Xinyang 464000, P.R. China}

\begin{abstract}
In the spatially flat case of loop quantum cosmology, the connection $\bar{k}$  is usually replaced by the holonomy $\frac{\sin(\bar{\mu}k)}{\bar{\mu}}$ in the effective theory. In this paper, instead of the standard $\bar{\mu}$ scheme, we use a generalised, undertermined  function $g(\bar{k},\bar{p})$
to represent the holonomy and by using the approach of anomaly free constraint algebra we fix all the counter terms in the constraints and find the restriction on the form of $g(\bar{k},\bar{p})$, then we derive the gauge invariant equations of motion of the scalar, tensor and vector perturbations and study the inflationary power spectra with generalised holonomy correction.
\end{abstract}

\maketitle

\section{Introduction}

  As an tentative theory, loop quantum cosmology (LQC) has made an enormous progress in applying the quantization scheme of loop quantum gravity (LQG)  to cosmological background variables in the past two decades. On the semi-classical level of  LQC, the quantization of geometric variables leads to several important quantum corrections to the classical theory, among them three quantum corrections are most often studied, namely the holonomy correction, the inverse volume correction and the quantum back-reaction \cite{Bojowald14a}. Roughly speaking, the holonomy correction comes from spatial discreteness, the inverse volume correction comes from a modified form of the inverse volume operator, the quantum back-reaction comes from coupling of quantum moments in the effective dynamics. So far, the explicit form of these corrections are far from unique and subjects to many quantization ambiguities. Let us take the holonomy correction for example, in many literatures the $SU(2)$ representation with minimal spin number $j=\frac{1}{2}$ is chosen, on the spatially flat background such a choice leads to the result that the classical connection variable $\bar{k}$ being replaced by the holonomy $\frac{\sin(\bar{\mu}k)}{\bar{\mu}}$ (where $\bar{\mu}$ depends on the scale factor). However, besides this standard $\bar{\mu}$ scheme, there are also other choices proposed to formulate the holonomy function, for instance, the representations with generic $j$ are studied in many works (see for example refs. \cite{Gaul01,Vandersloot05,Chiou09,Achour17}). Some other formulations involving the quantization ambiguities or higher powers of $\bar{\mu}$ can be found for instance in \cite{Hrycyna09,Yang09}.

  Besides the background dynamics, in order to find the footprints of LQC in the very early universe, it is necessary to study the cosmological perturbation theory. Currently, there are two main approaches often used in the LQC community to study the cosmological perturbations, which are separately called the `` dressed effective metric" approach and `` anomaly free constraint algebra" approach \footnote[1]{To be precise, there are actually three different approaches: the ``hybrid models" developed by Marug\'{a}n et al. , the ``dressed metric" developed by Ashtekar et al. and the ``anomaly free algebra" developed by Bojowald et al., however, the key ideas of the former two are quite similar, and for convenience of statement we put them together with a common name ``dressed effective metric".}. The philosophy of the two approaches are quite different, and so are their perturbation equations. Each approach has its own advantage and at the same time can also be questioned in some aspects. For the ``dressed effective metric " the reader is referred to \cite{Agullo12,M¨¦ndez12}. Here in this paper we only focus on the ``anomaly free constraint algebra" approach.

The key idea of the ``anomaly free constraint algebra" can be summarised as follows. In the canonical theory, the Hamiltonian of general relativity is described by three smeared constraints in terms of the Ashtekar variables, namely the Hamiltonian constraint, the diffeomorphsim constraint and the Gauss constraint. In the classical theory, the Poisson brackets between these constraints weakly vanish on the constraint surfaces and  form a first class constraint algebra. In the quantum theory where the constraint are quantum-modified, these constraints should also form a first class constraint algebra in order to make the theory consistent. When calculating the Poisson brackets, if we only consider the background dynamics, it is easy to check the quantum modified constraints still form a closed algebra, however, if we include perturbations into the constraints, the problem becomes quite subtle because at present we do not know how to include the explicit form of quantum modifications into the perturbed constraints, if we simply assume the perturbed constraints remain the classical form and calculate the Poisson bracket, the resulting algebra is no longer closed and some ``anomaly" terms appear. In order to remove these ``anomalies", a simple way is too add some counter terms which represent the quantum modifications into the perturbed constraints, then after fixing each counter term, a consistent effective theory is established.

The anomaly free constraint algebra approach was first systematically developed by Bojowald et al. in \cite{Bojowald08a,Bojowald08b} for inverse volume corrections. The results obtained in these papers were  used to study the scalar and tensor perturbations in \cite{Bojowald10,Li11}. Later, this method was applied to the scalar perturbations in the longitudinal gauge first in \cite{Wilson-Ewing12} for holonomy corrections in the standard $\bar{\mu}$ formulation and further developed in \cite{Cailleteau12a} without any gauge fixing. Moreover, the Hamilton-Jacobi approach of this method is studied in \cite{Cailleteau12b} and the consistency of the algebra with holonomy correction was investigated in \cite{Cailleteau12c}. The power spectra of different perturbations were studied in many works in recent years (see for example \cite{Linsefors13}).

In this paper, we use a generalised function $g(\bar{k},\bar{p})$ to represent the holonomy (where $\bar{p}$ stands for the conjugate momentum of $\bar{k}$), then with the method of anomaly free constraint algebra we attempt to find a consistent theory and study the cosmological perturbations. The structure of this paper is as follows, in section 2 we derive the anomaly free constraints and find the restrictions on $g(\bar{k},\bar{p})$, in section 3 we drive the scalar, vector and tensor perturbations. In section 4 we apply the results in section 3 to study the power spectra of the scalar and tensor perturbations in slow roll inflation, in section 5 we make some remarks.

\section{Anomaly free constraints} \label{Sperthol}
 For simplicity, in this paper we only focus on the linear perturbations in spatially flat case. In this section, we will fix the counter terms following the procedures formulated in \cite{Cailleteau12a}.

In the background Hamiltonian constraint,  the connection $\bar{k}$ is now replaced by a undetermined function $g(\bar{k},\bar{p})$ which consists of both $\bar{k}$ and its conjugate variable $\bar{p}$ and should satisfy the requirement that $g(\bar{k},\bar{p})\rightarrow\bar{k}$ when approaching the classical limit. The gravitational part of the effective background Hamiltonian constraint reads

\begin{equation}
H^{B}_g[N] =
\frac{1}{2 \kappa} \int_{\Sigma} d^3x \bar{N}
\left[ -6 \sqrt{\bar{p}}~g^2(\bar{k},\bar{p})\right],
\label{BHamilconstr
}
\end{equation}
where capital``B" stands for ``background" and $\bar{N}$ means the background part of the lapse function. The conjugate variables satisfy $\{\bar{k},\bar{p}\}=\frac{\kappa}{3V_0}$ where $\kappa=8\pi G$, $\Sigma$ is the fiducial cell that we do integration on and $V_0$ is its volume with respect to the background metric.

 When considering the linear perturbations,  the corresponding perturbed constraint should consist of quadratic terms of linear perturbed variables, after including the possible counter terms, the perturbed Hamiltonian constraint becomes \cite{Bojowald08a}
\begin{equation}
H^{P}_g[N] =\frac{1}{2 \kappa} \int_{\Sigma} d^3x \left[\delta N \mathcal{H}^{(1)}_g+\bar{N}\mathcal{H}^{(2)}_g\right],
\label{clHamilconstr}
\end{equation}
where
\begin{eqnarray}
\mathcal{H}^{(1)}_g&=&
-4\sqrt{\bar{p}}~(\bar{k}+\alpha_1)\delta^a_i\delta K^i_a
-\frac{1}{\sqrt{\bar{p}}}~(\bar{k}^2+\alpha_2)\delta^i_a\delta E^a_i\nonumber  \\
&&+\frac{2}{\sqrt{\bar{p}}}~(1+\alpha_3)\partial_a \partial^i \delta E^a_i ,  \nonumber  \\
\mathcal{H}^{(2)}_g &=&
\sqrt{\bar{p}}~(1+\alpha_4)\delta^a_k \delta^b_j\delta K^j_a \delta K^k_b
-\sqrt{\bar{p}}~(1+\alpha_5) (\delta^a_i\delta K^i_a )^2 \nonumber  \\
&&-\frac{2}{\sqrt{\bar{p}}}~(\bar{k}+\alpha_6) \delta E^a_i \delta K^i_a
-\frac{1}{2\bar{p}^{\frac{3}{2}}}~(\bar{k}^2+\alpha_7)\delta E^a_j \delta E^b_k \delta^k_a\delta^j_b \nonumber \\
&&+\frac{1}{4\bar{p}^{\frac{3}{2}}}(\bar{k}^2+\alpha_8) (\delta E^a_i \delta^i_a)^2
+\frac{1}{\bar{p}^{\frac{3}{2}}}(1+\alpha_9)Z^{cidj}_{ab}~( \partial_c \delta E^a_i) (\partial_d \delta E^b_j) , \nonumber
\end{eqnarray}
where $\delta K^i_a$ and $\delta E^a_i$ are the perturbed connection and perturbed densitized triad and satisfy the Poisson bracket $\{\delta K^i_a(x),\delta E^b_j(y)\}=\kappa \delta^i_j\delta^b_a\delta^3(x-y)$. The counter terms $\alpha_1$ to $\alpha_9$  are assumed as functions only of $\bar{k}$ and $\bar{p}$ in this paper. The $Z^{cidj}_{ab}$ in $\mathcal{H}^{(2)}_g$ is defined by \cite{Cailleteau12c}
 \begin{eqnarray}
 Z^{cidj}_{ab}:=\frac{1}{4} \epsilon_k^{~ef} \epsilon^k_{~mn} X^{mjd}_{be} X^{nic}_{af}  - \epsilon_k^{~ie} X^{kjd}_{be} \delta^c_a - \epsilon_k^{~ci} X^{kjd}_{ba}+ \frac{1}{2}\epsilon_k^{~ce} X^{kjd}_{be}\delta^i_a .\label{defZ}
 \end{eqnarray}
where
$X^{ijb}_{ca} :=  \epsilon_c^{~ij}\delta ^b_a - \epsilon_c^{~ib}\delta ^j_a + \epsilon^{ijb} \delta_{ca} + \epsilon_a^{~ib} \delta ^j_c$~.
Thus the holonomy corrected gravitational Hamiltonian constraint can be written as
\begin{equation}
 H_g^Q[N]=H_g^B[N]+H_g^P[N] .
 \end{equation}
In addition to the Hamiltonian constraint, there are also diffemorphism and Gauss constraints. Note that in LQG both constraints do not receive quantum corrections, if LQC is regarded as the correct cosmological reduced  model of LQG, then both constraints should keep the classical form on the effective level. In the spatially flat case we consider here, both constraints vanish on the background and only the perturbed parts exist, thus the linearly perturbed gravitational diffemorphism constraint and Gauss constraint can be written separately as
\begin{equation}
D_g[N^a] = \frac{1}{\kappa} \int_{\Sigma} d^3x \delta N^a \left[ \bar{p}\partial_a(\delta^b_i \delta K^i_b )
-\bar{p}(\partial_i \delta K^i_a)
-\bar{k}\delta^i_a (\partial_b \delta E^b_i )\right],
\label{diffconstr}
\end{equation}
and
\begin{equation}
G[\Lambda] = \frac{1}{\kappa} \int_{\Sigma} d^3x \delta\Lambda^i( \epsilon^{~~a}_{ij}\bar{p}~\delta K^j_a
+\epsilon^{~~k}_{ia}\bar{k}~\delta E^a_k ).
\label{guassconstr}
\end{equation}

Besides the gravitational part,  we introduce the minimally coupled scalar field as the matter field. As shown in \cite{Cailleteau12a}, the introduction of scalar matter field not only completes the theory but also helps to fix some of the counter terms. The total holonomy corrected Hamiltonian including both the background and perturbation of the scalar field can be expressed as
\begin{eqnarray}
H^Q_m[N] &:=&H_m^B[N]+H_m^P[N]\nonumber\\
 &:=&\int_{\Sigma} d^3x \bar{N}\mathcal{H}^{(0)}_m
 +\int_{\Sigma} d^3x \left[\delta N\mathcal{H}^{(1)}_m+\bar{N} \mathcal{H}^{(2)}_m\right] ,
\label{totmHamil}
\end{eqnarray}
where $\mathcal{H}^{(0)}_m$, $\mathcal{H}^{(1)}_m$, $\mathcal{H}^{(2)}_m$ are expressed as
\begin{eqnarray}
\mathcal{H}^{(0)}_m&=&\frac{\bar{\pi}^2}{2\bar{p}^{\frac{3}{2}}}
+\bar{p}^{\frac{3}{2}} V(\bar{\varphi}) ,\\
\mathcal{H}^{(1)}_m&=&\frac{\bar{\pi} }{\bar{p}^{\frac{3}{2}}}(1+\beta_1)\delta \pi-\frac{\bar{\pi}^2}{4\bar{p}^{\frac{5}{2}}}(1+\beta_2)~\delta^i_a \delta E^a_i \nonumber \\
&+&\bar{p}^{\frac{3}{2}} V_{,\varphi}(\bar{\varphi}) (1+\beta_3)\delta \varphi+\frac{\sqrt{\bar{p}}}{2}V(\bar{\varphi})(1+\beta_4)
\delta^i_a\delta E^a_i ,\\
\mathcal{H}^{(2)}_m&=& \frac{1}{2\bar{p}^{\frac{3}{2}}}(1+\beta_5)(\delta \pi)^2-\frac{\bar{\pi} }{2\bar{p}^{\frac{5}{2}}}(1+\beta_6)\delta \pi\delta^i_a \delta E^a_i
\nonumber\\
&+&\frac{1}{2}\bar{p}^{\frac{3}{2}} V_{,\varphi\varphi}(\bar{\varphi})(1+\beta_7) (\delta \varphi)^2
+\frac{\sqrt{\bar{p}}}{2} V_{,\varphi}(\bar{\varphi})(1+\beta_8)\delta \varphi\delta^i_a \delta E^a_i \nonumber \\
&+&\frac{\bar{\pi}^2}{16\bar{p}^{\frac{7}{2}}} (1+\beta_9)(\delta^i_a \delta E^a_i )^2+\frac{\bar{\pi}^2}{8\bar{p}^{\frac{7}{2}}}(1+\beta_{10})\delta^i_a \delta^j_b \delta E^a_j \delta E^b_i\nonumber\\
&+&\frac{V(\bar{\varphi})}{8\sqrt{\bar{p}}} (1+\beta_{11})(\delta^i_a \delta E^a_i )^2
-\frac{V(\bar{\varphi})}{4\sqrt{\bar{p}}}(1+\beta_{12})\delta^i_a \delta^j_b \delta E^a_j \delta E^b_i \nonumber\\
&+&\frac{1}{2} \sqrt{\bar{p}}(1+\beta_{13})\delta^{ab} (\partial_a \delta \varphi)  \partial_b \delta \varphi ,
\end{eqnarray}
where the conjugate variables satisfy $\{\bar{\varphi},\bar{\pi}\}=\frac{1}{V_0}$, $\{\delta\varphi(x),\delta\pi(y)\}=\delta^3(x-y)$, and $V_{,\varphi}:=\frac{d V(\varphi)}{d
\varphi}$, $V_{,\varphi\varphi}:=\frac{d^2 V(\varphi)}{d
\varphi^2}$ where $V$ is the potential. $\beta_1$ to $\beta_{13}$ are counter terms that vanish in classical limit and likewise are assumed as functions just of $\bar{k}$ and $\bar{p}$. Note that the background Hamiltonian constraint is the same as the classical one, which means the holonomy corrections do not affect the background Hamiltonian of the matter field.
Moreover, diffeomorphism constraint of the scalar matter reads simply
\begin{equation}
D_m[N^a] = \int_{\Sigma}d^3x \delta N^a \bar{\pi} (\partial_a \delta \varphi) ,
\end{equation}
which remains unchanged just like the gravitational part.

The total Hamiltonian constraint and diffeomorphism constraint are expressed by
\begin{eqnarray}
H_{tot}[N] &=& H^Q_g[N]+H^Q_m[N],   \\
D_{tot}[N^a] &=& D_g[N^a]+D_m[N^a].
\end{eqnarray}
The Hamiltonian can be written as sum of all the constraints:
\begin{equation}
\mathbf{H }= H_{tot}[N]+D_{tot}[N^a]+G[\Lambda].
\end{equation}
As mentioned above, the Poisson brackets between the constraints should weakly vanish in order to make the theory consistent. In the following we will calculate all these brackets, the details are rather technical in the subsection below, the uninterested reader can well skip them and directly jump to subsection 2.2 for the conclusion.
\subsection{The Poisson brackets}
In this section, we will derive all the anomaly equations generated by the constraint algebra. Since the diffeomorphism and Gauss constraint remain the classical form, the Poisson brackets between them do not generate any anomalies, hence the possible anomalies can only arise between three brackets, which are $\{H_{tot},D_{tot}\}$,$\{H_{tot},H_{tot}\}$ and $\{H_{tot},G\}$.
Now we calculate each of them in turn.
\subsubsection{$\{H_{tot}[N],D_{tot}[N^a]\}$}
\ \\
Using the expression of $H_g^Q$ and $H_m^Q$, after some calculations, the result can be written as sum of some independent terms,
\begin{eqnarray}
&&\{H_{tot}[N],D_{tot}[N^a]\}\nonumber\\
&&=- H_{tot}[\delta N^a \partial_a \delta N]
+\int_{\Sigma}  d^3x \frac{\sqrt{\bar{p}}}{\kappa}(\delta N^a \partial_a \delta N) \mathcal{A}_1\nonumber\\
&&+\int_{\Sigma}  d^3x \frac{\sqrt{\bar{p}}}{\kappa} \bar N \delta N^c \partial_c (\delta^a_i\delta K^i_a) \mathcal{A}_2
+\int_{\Sigma}  d^3x \frac{\sqrt{\bar{p}}}{\kappa} \bar N (\delta N^a \partial_i \delta K^i_a) \mathcal{A}_3\nonumber\\
&&+\int_{\Sigma}  d^3x \frac{\bar N}{\kappa\sqrt{\bar{p}}}  (\delta N^i \partial_a \delta E^a_i) \mathcal{A}_4
+\int_{\Sigma}  d^3x \frac{\bar N}{\kappa\sqrt{\bar{p}}} \delta N^c \partial_c (\delta^i_a\delta E^a_i)  \mathcal{A}_5\nonumber\\
&&+\int_{\Sigma} d^3x ( \delta N^c \partial_c \delta N )\bar p^{\frac{3}{2}} V(\bar{\varphi}) \mathcal{A}_{6}
+\int_{\Sigma} d^3x (  \delta N^c \partial_c \delta N)\left( \frac{\bar \pi^2}{2  \bar p^{\frac{3}{2}}}  \right) \mathcal{A}_{7}\nonumber\\
&&+ \int_{\Sigma} d^3x \bar N \bar p^{\frac{3}{2}}  (\delta N^c\partial_c \delta \varphi ) V_{,\varphi}(\bar{\varphi}) \mathcal{A}_{8}
+ \int_{\Sigma} d^3x \frac{\bar N}{\bar p^{\frac{3}{2}}} (\delta N^c\partial_c \delta \pi ) \bar \pi \mathcal{A}_{9}\nonumber\\
&&+\int_{\Sigma}  d^3x \frac{\bar N\sqrt{\bar{p}}}{2} (\delta N^i \partial_a \delta E^a_i) V(\bar{\varphi}) \mathcal{A}_{10}
+\int_{\Sigma}  d^3x \frac{\bar N\sqrt{\bar{p}}}{2} \delta N^c \partial_c (\delta^i_a\delta E^a_i) V(\bar{\varphi}) \mathcal{A}_{11}\nonumber\\
&&+\int_{\Sigma}  d^3x \frac{\bar N}{12\bar{p}^{\frac{5}{2}}} (\delta N^i \partial_a \delta E^a_i) \bar{\pi}^2 \mathcal{A}_{12}
+\int_{\Sigma}  d^3x \frac{\bar N}{4\bar{p}^{\frac{5}{2}}} \delta N^c \partial_c (\delta^i_a\delta E^a_i) \bar{\pi}^2 \mathcal{A}_{13} ,\nonumber\\
\end{eqnarray}
where
\begin{eqnarray}
 \mathcal{A}_1=3\bar{k}^2-3g^2(\bar{k},\bar{p})+2\bar{k}\alpha_1
 +\alpha_2 ,\\
 \mathcal{A}_2=-\frac{\partial g^2(\bar{k},\bar{p})}{\partial \bar{k}}+2\bar{k}+\bar{k}\alpha_5+\alpha_6 ,\\
 \mathcal{A}_3=\frac{\partial g^2(\bar{k},\bar{p})}{\partial \bar{k}}-2\bar{k}-\bar{k}\alpha_4-\alpha_6 ,\\
 \mathcal{A}_4=-\frac{1}{2}g^2(\bar{k},\bar{p})+\frac{1}{2}\bar{k}^2
 +\bar{k}\alpha_6
 -\frac{1}{2}\alpha_7-\bar{p} \frac{\partial g^2(\bar{k},\bar{p}) }{\partial \bar{p}} ,\\
 \mathcal{A}_5=\frac{1}{2}\alpha_7-\frac{1}{2}\alpha_8 ,\\
 \mathcal{A}_{6}=-\beta_4 ,\\
 \mathcal{A}_{7}=\beta_2-2\beta_1 ,\\
 \mathcal{A}_{8}=-\beta_8 ,\\
 \mathcal{A}_{9}=\beta_6-\beta_5 ,\\
 \mathcal{A}_{10}=-\beta_{12} ,\\
 \mathcal{A}_{11}=\beta_{12}-\beta_{11} ,\\
 \mathcal{A}_{12}=3\beta_{10} ,\\
 \mathcal{A}_{13}=2\beta_{6}-\beta_{9}-\beta_{10} ,
\end{eqnarray}

$\mathcal{A}_{1}$ to $\mathcal{A}_{13}$ are anomalies that should vanish, which means we have to solve the equations
\begin{equation}
\mathcal{A}_i=0 ,\quad(i=1,2...13).\nonumber
\end{equation}
\subsubsection{$\{H_{tot}[N_1],H_{tot}[N_2]\}$}
\ \\
The computation of this Poisson bracket gives the following results,
\begin{eqnarray}
&&\{H_{tot}[N_1],H_{tot}[N_2]\}\nonumber\\
&&=(1+\alpha_3)(1+\alpha_5)D_g \left[ \frac{\bar{N}}{\bar{p}} \partial^a(\delta N_2 -\delta N_1)  \right]\nonumber\\
&&+(1+\beta_1)(1+\beta_{13})  D_m\left[ \frac{\bar N}{\bar p} \partial^a (\delta N_2-\delta N_1)\right]\nonumber\\
&&+\int_{\Sigma} d^3x \frac{\bar{N}}{\kappa}\partial^a(\delta N_2 -\delta N_1)(\partial_i \delta K^i_a)\mathcal{A}_{14}\nonumber\\
&&+\int_{\Sigma} d^3x\frac{\bar{N}}{\kappa \bar{p}} \partial^a(\delta N_2 -\delta N_1)\delta^i_a(\partial_c \delta E^c_i) \mathcal{A}_{15}\nonumber\\
&&+\int_{\Sigma} d^3x\frac{\bar{N}}{\kappa} (\delta N_2 -\delta N_1)(\delta^a_i  \delta K^i_a) \mathcal{A}_{16}\nonumber\\
&&+\int_{\Sigma} d^3x \frac{\bar{N}}{\kappa \bar{p}} (\delta N_2 -\delta N_1)(\delta^i_a \delta E^a_i) \mathcal{A}_{17}\nonumber\\
&&+\int_{\Sigma}d^3x \bar{N} (\delta N_2-\delta N_1)
(\partial_a\partial^j \delta E^a_j)\left( \frac{\bar{\pi}^2}{4\bar{p}^3}-\frac{V(\bar{\varphi})}{2} \right) \mathcal{A}_{18}\nonumber\\
&&+\int_{\Sigma}d^3x \frac{\bar{N}}{\bar{p}^2}(\delta N_2-\delta N_1)
(\delta^a_i\delta K^i_a) \frac{\bar{\pi}^2}{2} \mathcal{A}_{19}\nonumber\\
&&+\int_{\Sigma}d^3x \bar{N}\bar{p}(\delta N_2-\delta N_1)
(\delta^a_i\delta K^j_a) V(\bar{\varphi}) \mathcal{A}_{20}\nonumber\\
&&+\int_{\Sigma}d^3x \frac{\bar{N}}{\bar{p}^3}(\delta N_2-\delta N_1)
(\delta_a^i\delta E^a_i)\frac{ \bar{\pi}^2 }{4}\mathcal{A}_{21}\nonumber\\
&&+\int_{\Sigma}d^3x \bar{N}(\delta N_2-\delta N_1)
(\delta_a^i\delta E^a_i) \frac{V(\bar{\varphi})}{2} \mathcal{A}_{22}\nonumber\\
&&+\int_{\Sigma}d^3x \frac{\bar{N}}{\bar{p}^2}(\delta N_2-\delta N_1)
(\delta \pi)\bar{\pi}\mathcal{A}_{23}\nonumber\\
&&+\int_{\Sigma}d^3x \bar{N}\bar{p}(\delta N_2-\delta N_1)
(\delta \varphi)V_{,\varphi}(\bar{\varphi})\mathcal{A}_{24}\nonumber\\
&&+\int_{\Sigma}d^3x \kappa\bar{N}\bar{p}(\delta N_2-\delta N_1)
(\delta_a^i\delta E^a_i)V^2(\bar{\varphi}) \mathcal{A}_{25}\nonumber\\
&&+\int_{\Sigma}d^3x \frac{\bar{N}}{\bar{p}}(\delta N_2-\delta N_1)(\delta_a^i\delta E^a_i)\frac{\bar{\pi} V_{,\varphi}(\bar{\varphi})}{2} \mathcal{A}_{26}\nonumber\\
&&+\int_{\Sigma}d^3x \frac{\kappa\bar{N}}{\bar{p}^2}(\delta N_2-\delta N_1)
(\delta_a^i\delta E^a_i)\frac{\bar{\pi}^2 V(\bar{\varphi})}{8} \mathcal{A}_{27}\nonumber\\
&&+\int_{\Sigma}d^3x \frac{\kappa\bar{N}}{\bar{p}^5}(\delta N_2-\delta N_1)(\delta_a^i\delta E^a_i)\frac{\bar{\pi}^4}{16} \mathcal{A}_{28}\nonumber\\
&&+\int_{\Sigma}d^3x \bar{N}(\delta N_2-\delta N_1)
(\delta \varphi)\left(\bar{\pi} V_{,\varphi\varphi}(\bar{\varphi})\right)\mathcal{A}_{29}\nonumber\\
&&+\int_{\Sigma}d^3x \kappa\bar{N}\bar{p}^2(\delta N_2-\delta N_1)
(\delta\varphi)\left(\frac{V_{,\varphi}(\bar{\varphi})V(\bar{\varphi})}{2}
-\frac{V_{,\varphi}(\bar{\varphi})\bar{\pi}^2}{4\bar{p}^3}\right)
\mathcal{A}_{30}\nonumber\\
&&+\int_{\Sigma}d^3x\frac{ \kappa\bar{N}}{\bar{p}}(\delta N_2-\delta N_1)(\delta \pi)\left(\frac{V(\bar{\varphi})\bar{\pi}
}{2}-\frac{\bar{\pi}^3}{4\bar{p}^3}\right)
\mathcal{A}_{31}\nonumber\\
&&+\int_{\Sigma}d^3x \bar{N}(\delta N_2-\delta N_1)
(\delta \pi)V_{,\varphi}(\bar{\varphi})\mathcal{A}_{32} ,\nonumber\\
\end{eqnarray}
where
\begin{eqnarray}
 \mathcal{A}_{14}&=&(1+\alpha_3)(\alpha_5 - \alpha_4) ,\\
 \mathcal{A}_{15}&=&-\frac{1}{2}\frac{\partial g^2(\bar{k},\bar{p})}{\partial \bar{k}}(1+\alpha_3)+(1+\alpha_3)(\bar{k}+\alpha_6)
 +\bar{k}(1+\alpha_3)(1+\alpha_5)\nonumber  \\
 &&+\frac{\partial g^2(\bar{k},\bar{p})}{\partial \bar{k}}\bar{p}\frac{\partial \alpha_3}{\partial \bar{p}}-\frac{1}{2}g(\bar{k},\bar{p})\left(g(\bar{k},\bar{p})
 +4\bar{p}\frac{\partial}{\partial \bar{p}}g(\bar{k},\bar{p}) \right)\frac{\partial \alpha_3}{\partial \bar{k}}\nonumber  \\
 &&-(\bar{k}+\alpha_1)(1+\alpha_9) ,\\
 \mathcal{A}_{16}&=&\frac{\partial g^2(\bar{k},\bar{p})}{\partial \bar{k}}(\bar{k}+\alpha_1)+2\frac{\partial g^2(\bar{k},\bar{p})}{\partial \bar{k}}\bar{p}\frac{\partial \alpha_1}{\partial \bar{p}}-2(\bar{k}+\alpha_6)(\bar{k}+\alpha_1)\nonumber  \\
 &&-g(\bar{k},\bar{p})\left(g(\bar{k},\bar{p})
 +4\bar{p}\frac{\partial}{\partial \bar{p}}g(\bar{k},\bar{p}) \right)(1+\frac{\partial \alpha_1}{\partial \bar{k}})\nonumber\\
 &&+\frac{1}{2}(\bar{k}^2+\alpha_2)(2+3\alpha_5-\alpha_4) ,\\
 \mathcal{A}_{17}&=&\frac{1}{2}(\bar{k}+\alpha_6)
 (\bar{k}^2+\alpha_2)-\frac{1}{4}\frac{\partial g^2(\bar{k},\bar{p})}{\partial \bar{k}}(\bar{k}^2+\alpha_2)+\frac{1}{2}\frac{\partial g^2(\bar{k},\bar{p})}{\partial \bar{k}}\bar{p}\frac{\partial \alpha_2}{\partial \bar{p}}\nonumber  \\
 &&-\left(g^2(\bar{k},\bar{p})+2\bar{p}\frac{\partial}{\partial \bar{p}}g^2(\bar{k},\bar{p}) \right)(\frac{1}{2}\bar{k}+\frac{1}{4}\frac{\partial \alpha_2}{\partial \bar{k}})\nonumber  \\
 &&+(\bar{k}+\alpha_1)
 (\frac{1}{2}\bar{k}^2+\frac{3}{2}\alpha_8-\alpha_7) ,\\
 \mathcal{A}_{18}&=&\frac{\partial \alpha_3}{\partial \bar{k}} ,\\
 \mathcal{A}_{19}&=&-(1+\frac{\partial \alpha_1}{\partial \bar{k}})+(1+\beta_2)(1+\frac{3}{2}\alpha_5-\frac{1}{2}\alpha_4) ,\\
 \mathcal{A}_{20}&=&(1+\frac{\partial \alpha_1}{\partial \bar{k}})-(1+\beta_4)(1+\frac{3}{2}\alpha_5-\frac{1}{2}\alpha_4) ,\\
 \mathcal{A}_{21}&=&-\frac{1}{2}g^2(\bar{k},\bar{p})\frac{\partial \beta_2}{\partial \bar{k}}
 -(\bar{k}+\frac{1}{2}\frac{\partial \alpha_2}{\partial \bar{k}})+(\bar{k}+\alpha_6)(1+\beta_2)\nonumber\\
 &&+(\bar{k}+\alpha_1)(5+3\beta_9+2\beta_{10})-(5+5\beta_2
 -2\bar{p}\frac{\partial \beta_2}{\partial \bar{p}})\frac{1}{2}\frac{\partial g^2(\bar{k},\bar{p})}{\partial \bar{k}} ,\\
 \mathcal{A}_{22}&=&\frac{1}{2}g^2(\bar{k},\bar{p})\frac{\partial \beta_4}{\partial \bar{k}}
 +(\bar{k}+\frac{1}{2}\frac{\partial \alpha_2}{\partial \bar{k}})-(\bar{k}+\alpha_6)(1+\beta_4)\nonumber\\
 &&+(\bar{k}+\alpha_1)(1+3\beta_{11}-2\beta_{12})-(1+\beta_4
 +2\bar{p}\frac{\partial \beta_4}{\partial \bar{p}})\frac{1}{2}\frac{\partial g^2(\bar{k},\bar{p})}{\partial \bar{k}} ,\\
 \mathcal{A}_{23}&=&-3(\bar{k}+\alpha_1)(1+\beta_6)
 +\frac{1}{2}\frac{\partial g^2(\bar{k},\bar{p})}{\partial \bar{k}}(3+3\beta_1-2\bar{p}\frac{\partial \beta_1}{\partial \bar{p}})\nonumber\\
 &&+\frac{g^2(\bar{k},\bar{p})}{2}\frac{\partial \beta_1}{\partial \bar{k}} ,\\
 \mathcal{A}_{24}&=&3(\bar{k}+\alpha_1)(1+\beta_8)-\frac{1}{2}\frac{\partial g^2(\bar{k},\bar{p})}{\partial \bar{k}}
 \left(3(1+\beta_3)+2p\frac{\partial \beta_3}{\partial \bar{p}}\right)\nonumber\\
 &&+\frac{g^2(\bar{k},\bar{p})}{2}\frac{\partial \beta_3}{\partial \bar{k}} ,\\
 \mathcal{A}_{25}&=&-\frac{1}{4}\frac{\partial \beta_4}{\partial \bar{k}} ,\\
 \mathcal{A}_{26}&=&-2-\beta_2-\beta_4+(1+\beta_3)(1+\beta_6)
 +(1+\beta_1)(1+\beta_8) ,\\
 \mathcal{A}_{27}&=&\frac{\partial \beta_2}{\partial \bar{k}}+\frac{\partial \beta_4}{\partial \bar{k}} ,\\
 \mathcal{A}_{28}&=&-\frac{\partial \beta_2}{\partial \bar{k}} ,\\
 \mathcal{A}_{29}&=&-(1+\beta_3)+(1+\beta_1)(1+\beta_7) ,\\
 \mathcal{A}_{30}&=&-\frac{\partial \beta_3}{\partial \bar{k}} ,\\
 \mathcal{A}_{31}&=&-\frac{\partial \beta_1}{\partial \bar{k}} ,\\
 \mathcal{A}_{32}&=&1+\beta_1-(1+\beta_3)(1+\beta_5) ,
 \end{eqnarray}
 Similarly as above, we have to solve the anomaly equations
 \begin{eqnarray}
 \mathcal{A}_i=0,\quad(i=14,15...32)
 \end{eqnarray}
 to close the algebra.
 Note that in obtaining the anomaly $\mathcal{A}_{15}$, we used the following relationship
 \begin{eqnarray}
 \left\{\int_{\Sigma} d^3x \delta N(\delta^a_i\delta K^i_a),~\int_{\Sigma} d^3x \bar{N}Z^{cidj}_{ab}~( \partial_c \delta E^a_i) (\partial_d \delta E^b_j) \right\}\nonumber\\
 =\int_{\Sigma} d^3x \kappa\bar{N}\delta N  (\partial^i\partial_a \delta E^a_i) .\label{bracketZk}
 \end{eqnarray}
 which can be proved after direct calculations using the definition of $Z^{cidj}_{ab}$ in (\ref{defZ}).

 \subsubsection{$\{H_{tot}[N],G[\Lambda]\}$}
  \ \\
 Finally, let us calculate the Poisson bracket between the Hamiltonian and Gauss constraint.
 \begin{eqnarray}
&&\{H_{tot}[N],G[\Lambda]\}\nonumber\\
&&=\int_{\Sigma} d^3x \frac{\bar{N}\sqrt{\bar{p}}}{\kappa}\epsilon_{ij}^{~~a}\delta \Lambda^i\delta K_a^j\mathcal{A}_{32}+\int_{\Sigma} d^3x\frac{\bar{N}}{\kappa\sqrt{\bar{p}}}\epsilon_{ia}^{~~k}\delta \Lambda^i\delta E^a_k\mathcal{A}_{33},
\end{eqnarray}
where
\begin{eqnarray}
 \mathcal{A}_{33}=-\frac{\partial g^2(\bar{k},\bar{p})}{\partial \bar{k}}+2\bar{k}+\alpha_6+\bar{k}\alpha_4 ,\label{A32}\\
 \mathcal{A}_{34}=\frac{1}{2}g^2(\bar{k},\bar{p})+\bar{p}\frac{\partial g^2(\bar{k},\bar{p})}{\partial \bar{p}}-\frac{1}{2}\bar{k}^2+\frac{1}{2}\alpha_7
 -\bar{k}\alpha_6 \label{A33} .
 \end{eqnarray}
 Notice that $ \mathcal{A}_{33}=- \mathcal{A}_{3}$ and $ \mathcal{A}_{34}=- \mathcal{A}_{4}$, which means the Gauss constraint does not bring any new anomalies.
 \subsection{Solution of the anomaly equations}
  Now there are totally  22 counters terms, which are $\alpha_1$, $\alpha_2$$\cdot\cdot\cdot$$\alpha_9$ , $\beta_1$, $\beta_2$$\cdot\cdot\cdot$$\beta_{13}$. To fix the explicit form of these terms, we have to solve totally 34 anomaly equations. Some of these equations are not independent of each other. However, even if we remove all the non-independent ones, the number of the remaining equations is still more than the number of counter terms, which often implies there are no consistent solutions. Nevertheless, considering that we have not specified the explicit form of $g(\bar{k},\bar{p})$ , hence the number of the unknown quantities increase, making it possible to get consistent solutions. In the following, we will check this possibility in detail.

  Let us start with the simplest cases. The equations $\mathcal{A}_6=0$, $\mathcal{A}_8=0$, $\mathcal{A}_{10}=0$, $\mathcal{A}_{11}=0$,
   $\mathcal{A}_{12}=0$ imply $\beta_4=\beta_8=\beta_{10}=\beta_{11}=\beta_{12}=0$. Using these results, from equations $\mathcal{A}_{19}=0$, $\mathcal{A}_{20}=0$ and $\mathcal{A}_{7}=0$, we get $\beta_2=\beta_4=\beta_1=0$. Combined with equations $\mathcal{A}_{23}=0$, $\mathcal{A}_{24}=0$, $\mathcal{A}_{26}=0$ and $\mathcal{A}_{30}=0$, we get $\beta_3=\beta_6=0$, substituting it into $\mathcal{A}_9=0$, $\mathcal{A}_{13}=0$,  $\mathcal{A}_{29}=0$, we have $\beta_5=\beta_9=\beta_7=0$. Now we have $\beta_1=\beta_2=\beta_3=\cdots=\beta_{12}=0$. We can check that these solutions automatically satisfy the other equations $\mathcal{A}_{21}=0$, $\mathcal{A}_{22}=0$, $\mathcal{A}_{25}=0$, $\mathcal{A}_{27}=0$, $\mathcal{A}_{28}=0$, $\mathcal{A}_{31}=0$, $\mathcal{A}_{32}=0$.

  Using the results above, the equation $\mathcal{A}_{23}=0$ gives
  \begin{equation}
  \alpha_1=\frac{1}{2}\frac{\partial g^2(\bar{k},\bar{p})}{\partial \bar{k}}-\bar{k} ,
  \end{equation}
  and from $\mathcal{A}_1=0$ we can further derive
  \begin{equation}
  \alpha_2=3g^2(\bar{k},\bar{p})-\bar{k}^2-\bar{k}\frac{\partial g^2(\bar{k},\bar{p})}{\partial \bar{k}} ,
  \end{equation}
then from the equations $\mathcal{A}_{14}=0$, $\mathcal{A}_{19}=0$, $\mathcal{A}_{20}=0$ we get
\begin{equation}
  \alpha_4=\alpha_5=\frac{1}{2}\frac{\partial^2 g^2(\bar{k},\bar{p})}{\partial \bar{k}^2}-1~,
  \end{equation}
substituting them into the equations $\mathcal{A}_2=0$, $\mathcal{A}_3=0$, we have
   \begin{equation}
  \alpha_6=\frac{\partial g^2(\bar{k},\bar{p})}{\partial \bar{k}}-\frac{1}{2}\bar{k}\frac{\partial^2 g^2(\bar{k},\bar{p})}{\partial \bar{k}^2}-\bar{k}~.
  \end{equation}
  It is easy to check this solution is consistent with the equations $\mathcal{A}_{21}=0$, $\mathcal{A}_{22}=0$.

   Furthermore, equations $\mathcal{A}_4=0$ and $\mathcal{A}_5=0$ give
  \begin{equation}
  \alpha_7=\alpha_8=-g^2(\bar{k},\bar{p})-\bar{k}^2-2\bar{p}\frac{\partial g^2(\bar{k},\bar{p})}{\partial \bar{p}}+2\bar{k}\frac{\partial g^2(\bar{k},\bar{p})}{\partial \bar{k}}-\bar{k}^2\frac{\partial^2 g^2(\bar{k},\bar{p})}{\partial \bar{k}^2}.
  \end{equation}
  Now among all the $\alpha_i$ only $\alpha_3$ and $\alpha_9$ are left to be determined. From the equation $\mathcal{A}_{18}=\frac{\partial \alpha_3}{\partial \bar{k}}=0$, we find
  \begin{equation}
  \alpha_3=f(\bar{p})~,
  \end{equation}
  where $f(\bar{p})$ can be any arbitrary function of $\bar{p}$ which vanishes in the classical limit, it means this term cannot be determined by holonomy corrections alone. One may wonder whether it can be determined by introducing other corrections, for instance the inverse volume correction.  However, in the paper \cite{Cailleteau14} where both holonomy and inverse volume correction are taken into account, this term still remains an arbitrary function of $\bar{p}$. It seems that in the spatially flat case it is impossible to fix the explicit form of this term by the current anomaly free approach.  In our forthcoming paper \cite{Han18}, we will extend the anomaly free approach by including the positive spatial curvature and derive the explicit expression of this term, where it clearly shows that this term should be zero in the spatially flat limit ($l_0\rightarrow 0$).
  Thus in the following we set $\alpha_3=0$.

  Substituting the expression of $\alpha_6$, $\alpha_5$, $\alpha_1$ into $\mathcal{A}_{15}=0$, we have
  \begin{equation}
  \alpha_9=\alpha_3+2\bar{p}\frac{\partial \alpha_3}{\partial \bar{p}}=0 .
  \end{equation}
  Furthermore, from the Poisson bracket $\{H_{tot}[N_1],H_{tot}[N_2]\}$, it is easy to see that the prefactor before $D_g$ and $D_m$ should be the same, hence
  \begin{equation}
  (1+\beta_1)(1+\beta_{13})=(1+\alpha_3)(1+\alpha_5) ,
  \end{equation}
  from which we get
  \begin{equation}
  \beta_{13}=\alpha_3+\alpha_5+\alpha_3\alpha_5=
  \frac{1}{2}\frac{\partial^2 g^2(\bar{k},\bar{p})}{\partial \bar{k}^2}-1 .
  \end{equation}
  So far we have derived the expressions of all the anomalies. However,  notice that we have not checked whether the anomalies $\mathcal{A}_{16}=0$  and $\mathcal{A}_{17}=0$ are satisfied. Substituting all the relevant counter terms into both equations, we find separately
  \begin{eqnarray}
  -\left(\frac{\partial g^2(\bar{k},\bar{p})}{\partial \bar{k}}\right)\frac{\partial }{\partial \bar{k}}\left(2g^2(\bar{k},\bar{p})-\bar{k}\frac{\partial g^2(\bar{k},\bar{p})}{\partial \bar{k}}-2\bar{p}\frac{\partial g^2(\bar{k},\bar{p})}{\partial \bar{p}}\right)\nonumber\\
  +\left(\frac{\partial^2 g^2(\bar{k},\bar{p})}{\partial \bar{k}^2}\right)\left(2g^2(\bar{k},\bar{p})-\bar{k}\frac{\partial g^2(\bar{k},\bar{p})}{\partial \bar{k}}-2\bar{p}\frac{\partial g^2(\bar{k},\bar{p})}{\partial \bar{p}}\right)=0~.
  \end{eqnarray}
  and
  \begin{eqnarray}
  \left(\bar{k}\frac{\partial g^2(\bar{k},\bar{p})}{\partial \bar{k}}\right)\frac{\partial }{\partial \bar{k}}\left(2g^2(\bar{k},\bar{p})-\bar{k}\frac{\partial g^2(\bar{k},\bar{p})}{\partial \bar{k}}-2\bar{p}\frac{\partial g^2(\bar{k},\bar{p})}{\partial \bar{p}}\right)\nonumber\\
  +\left(\bar{k}\frac{\partial^2 g^2(\bar{k},\bar{p})}{\partial \bar{k}^2}\right)\left(2g^2(\bar{k},\bar{p})-\bar{k}\frac{\partial g^2(\bar{k},\bar{p})}{\partial \bar{k}}-2\bar{p}\frac{\partial g^2(\bar{k},\bar{p})}{\partial \bar{p}}\right)=0 ,
  \end{eqnarray}
  from which it is not difficult to see that the equation
  \begin{eqnarray}
    2g^2(\bar{k},\bar{p})-\bar{k}\frac{\partial g^2(\bar{k},\bar{p})}{\partial \bar{k}}-2\bar{p}\frac{\partial g^2(\bar{k},\bar{p})}{\partial \bar{p}}=0
    \end{eqnarray}
  must hold,  which can be reduced to
    \begin{eqnarray}
    g(\bar{k},\bar{p})-\bar{k}\frac{\partial g(\bar{k},\bar{p})}{\partial \bar{k}}-2\bar{p}\frac{\partial g(\bar{k},\bar{p})}{\partial \bar{p}}=0.\label{restrg}
    \end{eqnarray}
      Solving this equation leads to the following solution
    \begin{eqnarray}
    g(\bar{k},\bar{p})= \bar{p}^\frac{1}{2}f(\frac{\bar{k}}{\bar{p}^\frac{1}{2}})~,
    \label{gform}
    \end{eqnarray}
    where $f(\frac{\bar{k}}{\bar{p}^\frac{1}{2}})$ is an arbitrary function of $\frac{\bar{k}}{\bar{p}^\frac{1}{2}}$ ( $\frac{\bar{k}}{\bar{p}^\frac{1}{2}}$ is treated as a whole variable here).  Moreover, since in the large scale limit $g(\bar{k},\bar{p})\rightarrow\bar{k}$, thus $f(\frac{\bar{k}}{\bar{p}^\frac{1}{2}})$ should equal $\frac{\bar{k}}{\bar{p}^\frac{1}{2}}$ in this limit.

    In the standard $\bar{\mu}$ scheme, we have
$g(\bar{k},\bar{p})=\frac{\sin(\bar{\mu}k)}{\bar{\mu}}$ where $\bar{\mu}=\sqrt{\frac{\Delta l^2_{pl}}{\bar{p}}}$ and the constant $\sqrt{\Delta}\sim \mathcal{O}(1)$, thus
\begin{eqnarray}
 f(\frac{\bar{k}}{\bar{p}^{\frac{1}{2}}})=\frac{1}{\sqrt{\Delta l^2_{pl}}}\sin(\sqrt{\Delta l^2_{pl}}\frac{\bar{k}}{\bar{p}^\frac{1}{2}}).
\end{eqnarray}
  Interestingly, we have confirmed the result in \cite{Cailleteau12a} from a more general perspective.

 Finally,  let us summarize the algebra between all the constraints,
    \begin{eqnarray}
    \left\{G[\Lambda],G[\Lambda']\right\} = 0 ,\\ \left\{D_{tot}[N^a],G[\Lambda]\right\} = 0 ,\\
    \left\{H_{tot}[N],G[\Lambda]\right\} = 0 ,\\
\left\{ D_{tot}[N^a_1],D_{tot} [N^a_2] \right\} =0 , \\
\left\{ H_{tot}[N],D_{tot}[N^a] \right\} = - H_{tot}[\delta N^a \partial_a \delta N] , \\
\left\{ H_{tot}[N_1],H_{tot}[N_2] \right\} = \left(\frac{1}{2}\frac{\partial^2 g^2(\bar{k},\bar{p})}{\partial \bar{k}^2}\right)D_{tot} \left[  \frac{\bar{N}}{\bar{p}} \partial^a(\delta N_2 - \delta N_1)\right] ,
\label{HHbrac}
\end{eqnarray}
 Obviously the prefactor $\frac{1}{2}\frac{\partial^2 g^2(\bar{k},\bar{p})}{\partial \bar{k}^2}$ equals 1 in the classical limit. If the function $f^2(\frac{\bar{k}}{\bar{p}^\frac{1}{2}})$ has a maximum at some phase space point, then $\frac{1}{2}\frac{\partial^2 g^2(\bar{k},\bar{p})}{\partial \bar{k}^2}$ will be negative near this point, which means the spacetime signature will become Euclidean for the perturbations and this is exactly what happens in the standard $\bar{\mu}$ scheme near the bounce.

 The phenomenon of signature change for holonomy corrections is one of the most important results of the anomaly free algebra approach. It was first found in \cite{Cailleteau12a} for scalar perturbations in the standard $\bar{\mu}$ scheme using this approach. Its unexpected consequences and deep implications were investigated in details in homogeneous LQC in \cite{Mielczarek12,Barrau16} and generally reviewed in \cite{Wilson-Ewing16}. It is clearly shown in these references that effective signature change around the bounce point could very possibly bring some observational effects under certain initial conditions and thus provides a new way to test the predictions of LQC. The signature change is also widely investigated in many other models with different symmetries, such as the spherical case \cite{Bojowald15a}, Gowdy cosmology \cite{Bojowald15b} and two dimensional covariant models \cite{Bojowald16a}. In this article, what we have done in the above actually shows that the signature change is a universal feature in the general $\bar{\mu}$ formulation.
 Moreover, it is worth mentioning that in LQG with self-dual variables, the effective constraint algebra remains unchanged from that of general relativity and the signature change does not appear. This disappearance has already been shown in spherically symmetric case \cite{Achour16a}, cosmological perturbations \cite{Achour16b} and Gowdy models, etc. \cite{Achour17a}. Further research is needed in order to verify whether the signature change really exists around the bounce point.
\section{Gauge invariant cosmological perturbations}
In this section, we will apply the previous results to study the linear cosmological perturbations. The calculations in this section are straightforward but quite tedious. For brevity, at some steps we will skip the tedious derivations and directly give the final equations.
\subsection{Background equations of motion}
 In order to study the cosmological perturbations, it is necessary to derive the equations of motion of background variables first.
Before the derivation, it is worth remarking that according to the previous research such as in \cite{Bolliet15a,Bolliet15b} that under certain initial condition, the power spectra of the perturbations will receive exponential growth for small comoving scales, if the growth turns out too fast, it could no longer be regarded as small perturbations and will have non-negligible back reactions on the background equation. In order to avoid such complexity, in this article we simply assume the background equations of motion will not be affected by the linear perturbations without investigating which functions of $f(\frac{\bar{k}}{\bar{p}^{\frac{1}{2}}})$ and which models or initial conditions will violate this assumption.
To keep clarity, in the following we will use the notation
\begin{eqnarray}
  \mathcal{G}^{(1)}(\bar{k},\bar{p})&:=&\frac{1}{2}\frac{\partial g^2(\bar{k},\bar{p})}{\partial \bar{k}}=\frac{\bar{p}}{2}\frac{\partial f^2(\frac{\bar{k}}{\bar{p}^{\frac{1}{2}}})}{\partial \bar{k}},\quad\quad
  \mathcal{G}^{(2)}(\bar{k},\bar{p}):=\frac{\partial \mathcal{G}^{(1)}(\bar{k},\bar{p})}{\partial \bar{k}},\nonumber\\
  \mathcal{G}^{(3)}(\bar{k},\bar{p})&:=&\frac{\partial \mathcal{G}^{(2)}(\bar{k},\bar{p})}{\partial \bar{k}}.
\end{eqnarray}
When approaching the classical limit, we have $\mathcal{G}^{(1)}\rightarrow \bar{k}$, $\mathcal{G}^{(2)}\rightarrow1$, $\mathcal{G}^{(3)}\rightarrow 0$.

By using the elementary Poisson brackets and the equation (\ref{restrg}), the Hamiltonian equations of motion for background variables are written as
 \begin{eqnarray}
 \dot{\bar{k}} &=&-\frac{3\bar{N}}{2\sqrt{\bar{p}}}g^2(\bar{k},\bar{p})
 +\frac{\bar{N}\bar{k}}{\sqrt{\bar{p}}} \mathcal{G}^{(1)}
+\frac{\kappa}{2} \sqrt{\bar{p}}\bar{N}
\left[-\frac{\bar{\pi}^2}{2\bar{p}^3}+V(\bar{\varphi}) \right] , \label{dotk}    \\
\dot{\bar{p}} &=&  2\bar{N} \sqrt{\bar{p}}\mathcal{G}^{(1)} , \label{dotp} \\
\dot{\bar{\varphi}} &=&   \bar{N} \frac{\bar{\pi}}{\bar{p}^{\frac{3}{2}}}~,  \label{dotvarphi} \\
\dot{\bar{\pi}} &=&  -\bar{N}  \bar{p}^{\frac{3}{2}} V_{, \varphi}(\bar{\varphi}) , \label{dotpi}
\end{eqnarray}
and the background Hamiltonian constraint gives
\begin{equation}
 \frac{3}{\kappa}\sqrt{\bar{p}}g^2(\bar{k},\bar{p})=\rho ,\label{bHconstr}
\end{equation}
where $\rho:=\frac{\bar{\pi}^2}{2\bar{p}^3}+V(\bar{\varphi})$.

It is known that different choice of the lapse function corresponds to using different time coordinate, when setting $\bar{N}=1$, we obtain equations with respect to the cosmic time $t$ and when $\bar{N}=\sqrt{\bar{p}}$ we obtain equations with respect to the conformal time $\eta$, thus $dt=\sqrt{\bar{p}}d\eta$.
Using equations (\ref{dotp}) and (\ref{bHconstr}), the Friedmann equation can be separately given in terms of the cosmic time and conformal time by
\begin{eqnarray}
H^2&=&\frac{\kappa}{3}\rho(\frac{\partial g(\bar{k},\bar{p})}{\partial \bar{k}})^2 ,\label{Friedcos}
\end{eqnarray}
or
\begin{eqnarray}
\mathcal{H}^2&=&\frac{\kappa \bar{p}}{3}\rho(\frac{\partial g(\bar{k},\bar{p})}{\partial \bar{k}})^2 ,\label{Friedcon}
\end{eqnarray}
where $H:=\frac{d\bar{p}}{2\bar{p}dt}$ and $\mathcal{H}:=\frac{d\bar{p}}{2\bar{p}d\eta}$ is separately called the Hubble parameter and conformal Hubble parameter. Note that $H=\frac{\mathcal{G}^{(1)}}{\bar{p}^{\frac{1}{2}}}$, $\mathcal{H}=\mathcal{G}^{(1)}$ and in the classical limit we have $H=\frac{\bar{k}}{\bar{p}^{\frac{1}{2}}}$ and $\mathcal{H}=\bar{k}$. For brevity, in the following we will abuse the notation by defining $\cdot=\frac{d}{dt}$ and $'=\frac{d}{d\eta}$.

The Klein-Gorden equation can be expressed by
\begin{equation}
\ddot{\bar{\varphi}}+3H\dot{\bar{\varphi}}+ V_{, \varphi}(\bar{\varphi})=0 ,\label{KLcos}
\end{equation}
or
\begin{equation}
\bar{\varphi}''+2\mathcal{H}\bar{\varphi}'+ \bar{p}V_{, \varphi}(\bar{\varphi})=0 .\label{KLcon}
\end{equation}
Finally, by taking the time derivative of equation (\ref{Friedcos}) and using (\ref{dotk}),(\ref{dotp}),(\ref{dotvarphi}),(\ref{dotpi}),  we derive the Raychaudhuri equation:
\begin{equation}
\dot{H}=-\frac{\kappa}{2}\mathcal{G}^{(2)}\dot{\bar{\varphi}}^2 .\label{Raycos}
\end{equation}
or
\begin{equation}
\mathcal{H}'=\mathcal{H}^2-\frac{\kappa}{2}\mathcal{G}^{(2)}\bar{\varphi}'^2 \label{Raycon}.
\end{equation}
\subsection{Gauge invariant perturbations}
Now we proceed to derive the equations of motion of the perturbed variables, by directly doing the Poisson brackets and using (\ref{restrg}), the canonical equations are given by
\begin{eqnarray}
\dot{\delta K^i_a}&=&\bar{k} (\partial_a \delta N^i)+\frac{2}{\sqrt{\bar{p}}} (\partial_a \partial^i \delta N)+\frac{\bar{N}\delta K_a^i}{\sqrt{p}} \left(\mathcal{G}^{(2)}\bar{k}-2\mathcal{G}^{(1)}\right)\nonumber \\
&-&\frac{\bar{N}\delta^{ij}}{2 \bar{p}^\frac{3}{2}} (Z^{cidj}_{ab}\partial_c \partial_d \delta E^b_j+Z^{cjdi}_{ba}\partial_d \partial_c \delta E^b_j)\nonumber \\
&+&\frac{\delta^i_a\delta N}{2\sqrt{\bar{p}}}\left( 2\mathcal{G}^{(1)}\bar{k}-3g^2(\bar{k},\bar{p})-\frac{\kappa\bar{\pi}^2}{2\bar{p}^2}
+\kappa \bar{p} V(\bar{\varphi})\right)\nonumber \\
&+&\frac{\bar{N}(\delta_a^j\delta_c^i\delta E^c_j)}{\bar{p}^{\frac{3}{2}}}\left(\frac{3}{2}g^2(\bar{k},\bar{p})
+\mathcal{G}^{(2)}\bar{k}^2-3\mathcal{G}^{(1)}\bar{k}
+\frac{\kappa\bar{\pi}^2}{\bar{p}^2}-\frac{\kappa \bar{p}V(\bar{\varphi})}{2}\right)\nonumber \\
&-&\frac{\bar{N}(\delta_c^j\delta E^c_j)}{2\bar{p}^{\frac{3}{2}}}\delta_a^i\left(\frac{3}{2}g^2(\bar{k},\bar{p})
+\mathcal{G}^{(2)}\bar{k}^2-3\mathcal{G}^{(1)}\bar{k}-\frac{\kappa\pi^2}{2\bar{p}^2}-\kappa \bar{p}V(\bar{\varphi})\right)\nonumber \\
&-&\frac{\kappa \bar{N} \delta_a^i}{2\bar{p}^{\frac{5}{2}}}\pi \delta\pi+\frac{\sqrt{\bar{p}}\kappa \bar{N} \delta_a^i}{2}V_{,\varphi}(\bar{\varphi})\delta \varphi , \label{dotdeltak}\\
\dot{\delta E^a_i}&=& -\bar{p} (\partial_i \delta N^a -\delta^a_i\partial_c \delta N^c)+2\delta N\delta^a_i\sqrt{\bar{p}}\mathcal{G}^{(1)}+\bar{N} \sqrt{\bar{p}}(\delta^c_j\delta K^j_c )\delta^a_i\mathcal{G}^{(2)} \nonumber\\
&-&\bar{N}\sqrt{\bar{p}}(\delta^c_i \delta^a_j\delta K^j_c )\mathcal{G}^{(2)}+\frac{N}{\sqrt{\bar{p}}}\left(2\mathcal{G}^{(1)}-\mathcal{G}^{(2)}\bar{k}\right)\delta E^a_i,\label{dotdeltaE}\\
\dot{\delta\varphi}&=&\frac{\pi\delta N}{\bar{p}^{\frac{3}{2}}}+\frac{\bar{N}\delta \pi }{\bar{p}^{\frac{3}{2}}}-\frac{\bar{N}\bar{\pi}}{\bar{p}^{\frac{5}{2}}}(\delta_a^i\delta E^a_i) ,\label{dotdeltaphi}\\
\dot{\delta \pi}&=&\bar{\pi}\partial_a\delta N^a-\bar{p}^{3/2} V_{,\varphi}(\bar{\varphi})  \delta N-\bar{N}\bar{p}^{\frac{3}{2}}V_{,\varphi\varphi}(\bar{\varphi})\delta\varphi
-\frac{\bar{N}\bar{p}^{\frac{1}{2}}}{2}V_{,\varphi}(\bar{\varphi})\delta_a^i\delta E^a_i\nonumber\\
&+&\bar{N}\sqrt{\bar{p}}(\partial^a\partial_a\delta \varphi)\mathcal{G}^{(2)} .\label{dotdeltapi}
\end{eqnarray}
To use the above equations to study the gauge invariant perturbations, we decompose the perturbed densitized triad  into the scalar, vector and tensor modes
\begin{eqnarray}
\delta E^a_i&=&\bar{p}\Big[-2\psi\delta^a_i+(\delta^a_i\partial^d\partial_d-\partial^a \partial_i)E-a_1\partial^aF_i-a_2\partial_iF^a\nonumber\\
&&-\frac{1}{2}b_1h^a_{~i}-\frac{1}{2}b_2h^{~a}_{i}\Big] ,
\end{eqnarray}
where $a_1$, $a_2$ and $b_1$, $b_2$ are arbitrary real numbers satisfying $a_1+a_2=1$, $b_1+b_2=1$. Moreover, the perturbed lapse function and shift vector can be decomposed as
\begin{eqnarray}
\delta N=\bar{N}\phi ,\quad\quad\quad\delta N^a=\partial^a B+S^a ,
\end{eqnarray}
where $\partial_a F^a=\partial^i F_i=\partial_a S^a=0$, $\partial_a h^a_{~i}=\partial^i h^a_{~i}=0$ and $h^i_{~i}=0$. Since we only focus on the gauge invariant part of the perturbed variables, i.e. the physical parts which are invariant both under internal (Gauss) transformation and external (diffeomorphism) transformation, firstly, by doing Poisson brackets with the Gauss constraint, we easily find that the internally gauge invariant part of $\delta E^a_i$ should be its symmetric part, i.e. the one satisfies $a_1=a_2=\frac{1}{2}$, $b_1=b_2=\frac{1}{2}$ and is defined by
\begin{eqnarray}
\delta E^{(a}_{~i)}=\bar{p}\Big[-2\psi\delta^a_i
+(\delta^a_i\partial^d\partial_d- \partial^a\partial_i)E-\frac{1}{2}\partial^aF_i
-\frac{1}{2}\partial_iF^a-\frac{1}{2} h^{(a}_{~i)}\Big] ,\label{ginvE}
\end{eqnarray}
 where $h^{(a}_{~i)}:=\frac{1}{2}(h^a_{~i}+h^{~a}_{i})$.

 Substituting (\ref{ginvE}) into the equation (\ref{dotdeltaE}), using $\mathcal{G}^{(1)}=\mathcal{H}$, after short calculations, we find the corresponding internally gauge invariant $\delta K_{(a}^{~i)}$, which satisfies
\begin{eqnarray}
\mathcal{G}^{(2)}\delta K_{(a}^{~ i)}&=&\Big[-\delta^a_i\left(\psi'+\bar{k}\mathcal{G}^{(2)}\psi
+\mathcal{H}\phi\right)
+\partial_a\partial^i\left[\bar{k}\mathcal{G}^{(2)}E
-(B-E')\right]\Big]\nonumber\\
&+&\frac{1}{2}\Big[\bar{k}\mathcal{G}^{(2)}
(\partial_aF^i+\partial^iF_a)
+(\partial_aF^i+\partial^iF_a)'
-(\partial_aS^i+\partial^iS_a)\Big]\nonumber\\
&+&\frac{1}{2}\Big[{h^{(i}_{~a)}}'
+\bar{k}\mathcal{G}^{(2)}h^{(i}_{~a)}\Big] .\label{decomdeltak}
\end{eqnarray}
Notice that what we has done above is actually to reduce the full phase space with ``$ 9\times\infty$" degrees of freedom to a reduced phase space with ``$6\times\infty$" degrees of freedom, upon which basic conjugate variables are $\delta E^{(a}_{~i)}$,
$\delta K_{(a}^{~ i)}$ with $\{\delta K_{(a}^{~ i)}(x),\delta E^{(c}_{~j)}(y) \}=\frac{\kappa}{2}(\delta_a^c\delta^i_j+\delta_{aj}\delta^{ci})\delta^3(x-y)$.

Now let us derive the diffeomorphism invariant quantities, according to \cite{Bojowald08b,Cailleteau12a}, consider a small coordinate transformation
\begin{eqnarray}
x^{\mu}\rightarrow x^{\mu}+\xi^{\mu}~,\quad\quad \xi^{\mu}:=(\xi^{0},\xi^{a})~,\quad \xi^a:=\partial^a\xi+\tilde{\xi}^a ,
\end{eqnarray}
where $\partial_a\tilde{\xi}^a=0$,
the infinitesimal gauge transformation of a phase space variable $X$ under such transformation is given by
\begin{eqnarray}
\delta_{[\xi^0, \xi^a]}X:=\{X, H^{P}_{tot}[\bar{N}\xi^0]+D_{tot}[\xi^a]\} ,\label{dtransX}
\end{eqnarray}
where
\begin{eqnarray}
H^{P}_{tot}[\bar{N}\xi^0]:=\frac{1}{2 \kappa} \int_{\Sigma} d^3x \bar{N}\xi^0 \mathcal{H}^{(1)}_g+\int_{\Sigma} d^3x \bar{N}\xi^0 \mathcal{H}^{(1)}_m .
\end{eqnarray}
From (\ref{dtransX}) and (\ref{HHbrac}), it is easy to derive the gauge transformation of $X$'s time derivative, which is given by
\begin{eqnarray}
\delta_{[\xi^0, \xi^a]}\dot{X}=\dot{(\delta_{[\xi^0, \xi^a]} X)}-\mathcal{G}^{(2)}\delta_{[0, \partial^a\xi^0]}X .\label{dtransdotx}
\end{eqnarray}
Now we separately study the different modes of perturbations. Since the scalar modes automatically commute with the Gauss constraint, we only need to consider the diffeomorphism invariant parts, using (\ref{dtransX}), (\ref{dtransdotx}) and (\ref{restrg}), it is not difficult to find the gauge invariant variables $\Phi$, $\Psi$, $\delta \tilde{\varphi}$ for scalar perturbations, which satisfy
\begin{eqnarray}
\mathcal{G}^{(2)}\Phi &=& \mathcal{G}^{(2)}\phi+(B'-E'') +\left(\mathcal{H}-\mathcal{G}^{(3)}\bar{k}'
+\mathcal{G}^{(3)}\mathcal{H}\bar{k}\right)(B-E') ,\label{Phi} \\
\mathcal{G}^{(2)}\Psi &=&\mathcal{G}^{(2)}\psi- \mathcal{H}(B-E'),\label{Psi} \\
\mathcal{G}^{(2)}\delta\tilde{\varphi}
&=&\mathcal{G}^{(2)}\delta \varphi+\bar{\varphi}'(B-E') .\label{givarphi}
\end{eqnarray}
It is easy to see that in the classical limit $\Phi$ and $\Psi$ become the familiar Bardeen potentials.

Substituting (\ref{decomdeltak}) into equation (\ref{dotdeltak}), using (\ref{Phi}) and (\ref{Psi}), we find the off-diagonal part of the equation (\ref{dotdeltak}) gives
\begin{equation}
\Psi=\Phi ,
\end{equation}
which is exactly the same as the classical theory.
 Using this relation as well as the the Hamiltonian and diffeomorphism constraints, after a huge amount of calculations, we find the diagonal part of (\ref{dotdeltak}) gives the equation of motion of $\Psi$:
\begin{eqnarray}
&&\Psi''
+2(\mathcal{H}-\frac{\bar{\varphi}''}{\bar{\varphi}'})\Psi'
+2(\mathcal{H}'
-\mathcal{H}\frac{\bar{\varphi}''}{\bar{\varphi}'})\Psi
+\frac{\mathcal{G}^{(3)}}{\mathcal{G}^{(2)}}
\left(-\bar{k}'+\mathcal{H}\bar{k}\right)(\Psi'+\mathcal{H}\Psi)\nonumber\\
&&-\mathcal{G}^{(2)}\nabla^2\Psi=0 .\label{eomPsi}
\end{eqnarray}
Obviously (\ref{eomPsi}) can reproduce the equations of motion for $\Psi$ in the classical limit. It is worth emphasizing that this equation is only valid when $\mathcal{G}^{(2)}\neq 0$, at first sight, one may multiply each term by the factor $\mathcal{G}^{(2)}$ to avoid this trouble, however, this will generate an additional restriction $\Psi'+\mathcal{H}\Psi=0$ at the point $\mathcal{G}^{(2)}=0$, which is unnatural.

 Inserting the definition (\ref{givarphi}) into equation (\ref{dotdeltaphi}) and (\ref{dotdeltapi}), we get the equation of motion of the gauge invariant perturb scalar field $\delta\bar{\varphi}$,
\begin{eqnarray}
 \delta \tilde{\varphi}''+2\mathcal{H}  \delta \tilde{\varphi}'
+\bar{p}V_{,\varphi\varphi}(\bar{\varphi} ) \delta\bar{\varphi}
+2 \bar{p} V_{,\varphi}(\bar{\varphi} ) \Psi -4\bar{\varphi}'\Psi'-\mathcal{G}^{(2)}\nabla^2 \delta \tilde{\varphi}=0 .\label{eomtildphi}
\end{eqnarray}
 Combining equations (\ref{eomPsi}) with (\ref{eomtildphi}), straightforward calculation gives the Mukhanov equation
\begin{eqnarray}
v_S''-\mathcal{G}^{(2)}\nabla^2v_S-\frac{z_S''}{z_S}v_S= 0~, \label{SMukheqn}
\end{eqnarray}
where $v_S:= \sqrt{\bar{p}} \left( \delta \tilde{\varphi} + \frac{\bar{\varphi}'}{\mathcal{H}} \Psi \right)$ is the Mukhanov variable and $z_S := \sqrt{\bar{p}} \frac{\bar{\varphi}'}{\mathcal{H}}$. Note that both $v_S$ and $z_S$ are not well defined at $\mathcal{H}=0$, what's more, since $\mathcal{H}=\mathcal{G}^{(1)}$, thus near the point $\mathcal{H}=0$, $\mathcal{G}^{(2)}$ becomes negative, the solution of $v_S$ is not stable from the perspective of perturbation theory.

 It is easy to check that all the scalar perturbation equations obtained above can reproduce the equations derived in \cite{Cailleteau12a,Cailleteau12c} once we set $g(\bar{k},\bar{p})=\frac{\sin(\bar{\mu}\bar{k})}{\bar{\mu}}$.

 Now we consider the gauge invariant vector perturbations, by using (\ref{dtransdotx}), it is easy to find that
\begin{eqnarray}
V^a:=S^a-{F^a}'
\end{eqnarray}
is gauge invariant under coordinate transformations. Since as shown above the variable invariant with respect to Gauss constraint should be symmetrized, thus we define
\begin{eqnarray}
\Sigma^a_i=\frac{1}{2}(\partial_iV^a+\partial^aV_i) ,
\end{eqnarray}
which is invariant both under internal and external gauge transformations. By using (\ref{dotdeltak}) again, the equation of motion for $\Sigma^a_i$ reads
\begin{eqnarray}
{\Sigma^a_i}'+2\mathcal{H}\Sigma^a_i
-\frac{\mathcal{G}^{(3)}}{\mathcal{G}^{(2)}}\left(\bar{k}'-\mathcal{H}\bar{k}\right)
\Sigma^a_i=0 .\label{eomSigma}
\end{eqnarray}
Similarly, this equation is also not well defined at $\mathcal{G}^{(2)}=0$. Moreover, we find that in the standard $\bar{\mu}$ scheme this equation is quite different from the vector perturbation equation (69) derived in \cite{Mielczarek12a}, for which we will give a short explanation in the last section.

Finally, since from (\ref{dtransX}) it is easy to see that the tensor perturbation is naturally diffeomorphism invariant, we only need to consider the internally invariant $h^{(a}_{~i)}$, its equation of motion reads
\begin{eqnarray}
{h^{(a}_{~i)}}''+2\mathcal{H}
{h^{(a}_{~i)}}'-\frac{\mathcal{G}^{(3)}}{\mathcal{G}^{(2)}}(\bar{k}'-
\mathcal{H}\bar{k}){h^{(a}_{~i)}}'-\mathcal{G}^{(2)}\nabla^2h^{(a}_{~i)}=0 .\label{eomh}
\end{eqnarray}
Defining
\begin{eqnarray}
z_T=\sqrt{\frac{p}{\mathcal{G}^{(2)}}},\quad\quad\quad v_T=\frac{z_T}{\sqrt{2\kappa}}h^{(a}_{~i)},
\end{eqnarray}
we can turn (\ref{eomh}) into the tensor Mukhanov equation
\begin{eqnarray}
{v_T}''-\mathcal{G}^{(2)}\nabla^2 v_T-\frac{{z_T}''}{z_T}v_T =0 . \label{TMukheqn}
\end{eqnarray}
\section{Power spectra with generalised holonomy corrections}
 In this section, we apply the results obtained in the last section to derive the holonomy corrected scalar and tensor power spectrum during the slow-roll inflation. Since in the slow-roll period, the energy density is much lower than the typical quantum gravity scale and thus can be regarded quite close to the classical limit, recall that $g(\bar{k},\bar{p})=\bar{p}^{\frac{1}{2}}f(\frac{\bar{k}}{\bar{p}^{\frac{1}{2}}})$, we assume the function $f(\frac{\bar{k}}{\bar{p}^{\frac{1}{2}}})$ can be expanded into the series
 \begin{eqnarray}
 f(\frac{\bar{k}}{\bar{p}^{\frac{1}{2}}})=c_1(\frac{\bar{k}}{\bar{p}^{\frac{1}{2}}})
 +c_2(\frac{\bar{k}}{\bar{p}^{\frac{1}{2}}})^2
 +c_3(\frac{\bar{k}}{\bar{p}^{\frac{1}{2}}})^3+\cdots
 +c_n(\frac{\bar{k}}{\bar{p}^{\frac{1}{2}}})^n+\cdots .\label{Tseries}
\end{eqnarray}

Note that in the slow roll period we require $g(\bar{k},\bar{p})\rightarrow \bar{k}$ , from which we conclude $c_1=1$ and all the other terms should be much smaller than the first one, hence the first term stands for the classical contribution and the higher order terms represent the generalised holonomy correction. The requirement that all the higher order terms remain much smaller compared to the first one during the slow roll inflation is not trivial because both $\bar{k}$ and $\bar{p}$ change rapidly with respect to the cosmic time in this period, but considering that near the classical limit we have  $\frac{\bar{k}}{\bar{p}^{\frac{1}{2}}}\simeq H$ and in the slow roll inflation the Hubble parameter changes very slowly, hence each term in the series remain almost as a constant during the inflation.

 In the standard $\bar{\mu}$ formulation, $f(\frac{\bar{k}}{\bar{p}^{\frac{1}{2}}})=\frac{1}{\alpha}\sin(\alpha\frac{\bar{k}}{\bar{p}^\frac{1}{2}})$
where the coefficient $\alpha\propto l_{pl}$, its Taylor series read
\begin{eqnarray}
 f(\frac{\bar{k}}{\bar{p}^{\frac{1}{2}}})=\frac{\bar{k}}{\bar{p}^{\frac{1}{2}}}
 -\frac{\alpha^2}{3~!}(\frac{\bar{k}}{\bar{p}^{\frac{1}{2}}})^3
 +\frac{\alpha^4}{5~!}(\frac{\bar{k}}{\bar{p}^{\frac{1}{2}}})^5+\cdots .
\end{eqnarray}
Among all the quantum correction terms the leading one is proportional to $(\frac{\bar{k}}{\bar{p}^{\frac{1}{2}}})^3$. Due to the extremely small value of $\alpha$ and considering the discussion above, it is justified to only keep the leading correction term and discard all the other higher order terms in slow-roll inflation. Motivated by this, we assume that in the slow-roll period the function $g(\bar{k},\bar{p})$ can be approximated by
\begin{eqnarray}
 g(\bar{k},\bar{p})\simeq\bar{p}^{\frac{1}{2}}\left(\frac{\bar{k}}{\bar{p}^{\frac{1}{2}}}
 +c_m(\frac{\bar{k}}{\bar{p}^{\frac{1}{2}}})^m\right)=\bar{k}\left(
 1+c_m(\frac{\bar{k}}{\bar{p}^\frac{1}{2}})^{m-1}\right) ,\label{gapprox}
\end{eqnarray}
where the integer $m$ denotes the leading correction determined by the explicit form of $f(\frac{\bar{k}}{\bar{p}^{\frac{1}{2}}})$. Let us define the dimensionless parameter
\begin{eqnarray}
\delta_Q=c_m(\frac{\bar{k}}{\bar{p}^\frac{1}{2}})^{m-1},
\end{eqnarray}
 where we ask $\delta_Q\ll 1$, considering it is slowly changing during the slow-roll period,  the Friedmann equation (\ref{Friedcos}) can be written as
\begin{eqnarray}
H^2=\frac{\kappa}{3}\rho\left(1+2m\delta_Q\right) ,\label{slFr}
\end{eqnarray}
which can be approximated by
\begin{eqnarray}
H^2=\frac{\kappa}{3}V(\varphi)\left(1+2m\delta_Q\right) ,\label{slrFr}
\end{eqnarray}
if the slow roll conditions are satisfied. The Raychaudhuri equation (\ref{Raycos}) becomes
\begin{eqnarray}
\dot{H}=-\frac{\kappa}{2}\dot{\bar{\varphi}}^2\Big(1+m(m+1)\delta_Q\Big) ,\label{Rayslr}
\end{eqnarray}
and the Klein-Gorden equation is approximated by
\begin{eqnarray}
\dot{\bar{\varphi}}= -\frac{V_{,\varphi}(\bar{\varphi})}{3H} ,\label{slrFr}
\end{eqnarray}
the slow roll parameters are defined as
\begin{eqnarray}
\epsilon=-\frac{\dot{H}}{H^2}=
\frac{3}{2}\frac{\dot{\bar{\varphi}^2}}{V(\bar{\varphi})}
\frac{\Big(1+m(m+1)\delta_Q\Big)}{\Big(1+2m\delta_Q\Big)}\simeq\epsilon_V
\Big(1+m(m-3)\delta_Q\Big) ,\label{slrpara1}\\
\delta=\frac{\ddot{\bar{\varphi}}}{H\dot{\bar{\varphi}}}\simeq\epsilon_V
\Big(1+m(m-3)\delta_Q\Big)-\eta_V\Big(1-2m\delta_Q\Big) ,\label{slrpara2}
\end{eqnarray}
where the shape of inflation potential $\epsilon_V:=\frac{1}{2\kappa}(\frac{V_{,\varphi}}{V})^2$,
$\eta_V:=\frac{1}{\kappa}\frac{V_{,\varphi\varphi}}{V}$. Ignoring the the quadratic terms of $\delta_Q$, from (\ref{slrpara1}), it is not difficult to find
\begin{eqnarray}
\mathcal{H}:=\sqrt{p}H\simeq-\frac{1}{\eta}(1+\epsilon) .\label{mhslr}
\end{eqnarray}
\subsection{Scalar power spectrum}
Using the slow roll parameters and the equation (\ref{mhslr}), after direct calculations, the coefficient $\frac{{z_S}''}{z_S}$  defined in (\ref{SMukheqn})  becomes
\begin{eqnarray}
\frac{{z_S}''}{z_S}&=&\frac{1}{\eta^2}(2+6\epsilon+3\delta)\nonumber\\
&\simeq&\frac{1}{\eta^2}\Big(2+9\epsilon_V\big(1+m(m-3)\delta_Q\big)-3\eta_V\big(1-2m\delta_Q\big)\Big)
 .\label{SMukcoeff}
\end{eqnarray}
Thus the the scalar Mukhanov equation can be written as
\begin{eqnarray}
{v_S}''-\Big(1+m(m+1)\delta_Q\Big)\nabla^2 v_S - \frac{1}{\eta^2}\Big({\nu_S}^2-\frac{1}{4}\Big) v_S = 0~,\label{slrSMukheqn}
\end{eqnarray}
where
\begin{eqnarray}
|\nu_S|=\frac{3}{2}+3\epsilon_V\Big(1+m(m-3)\delta_Q\Big)-\eta_V\Big(1-2m\delta_Q\Big)~.
\end{eqnarray}
In the momentum space, (\ref{slrSMukheqn}) becomes
\begin{eqnarray}
{v^S_k}''+\Big(1+m(m+1)\delta_Q\Big)k^2 v^S_k - \frac{1}{\tau^2}\Big({\nu_S}^2-\frac{1}{4}\Big) v^S_k = 0~,\label{slrkSMukheqn}
\end{eqnarray}
Since $\delta_Q\ll1$ and varies very slowly, the coefficient $(1+m(m+1)\delta_Q)$ can be roughly regarded as a constant and the approximate solution of (\ref{slrkSMukheqn}) is
\begin{eqnarray}
v^{S}_{k}(\eta)\simeq\sqrt{-\eta}\Big[a_kH_{\nu1}(-\xi k\eta)+b_kH_{\nu2}(-\xi k\eta)\Big]~,\label{approsl}
\end{eqnarray}
where $\xi:=\sqrt{\mathcal{G}^{(2)}}=\Big(1+\frac{m}{2}(m+1)\delta_Q\Big)$, $H_{\nu1}$ and $H_{\nu2}$ are separately the first and second class Hankel functions.

As we know, the choice of initial conditions plays a crucial role in deriving the power spectrum, different choices of initial states may lead to completely different form of the power spectrum. In loop quantum cosmology, various initial conditions for the scalar and tensor perturbations have been considered \cite{Bolliet15a,Mielczarek14a,Mielczarek14b,Chen16}. In this paper, for simplicity, the initial states are chosen at the point right before the onset of inflation where $|\xi k\tau|\gg1$, thus the last term in (\ref{slrkSMukheqn}) can be neglected and
\begin{eqnarray}
v^{S}_{k}(\eta)\simeq\frac{e^{-ik\xi \eta}}{\sqrt{2\xi k}}, \label{vinitial}
\end{eqnarray}
  i.e. only the incoming modes are chosen, which is similar as in the classical theory. From the asymptotic behavior of the Hankel function
\begin{eqnarray}
H_{\nu1}(-\xi k\eta\gg1)\simeq \sqrt{\frac{2}{-\xi k\pi\eta }}e^{i(-\xi k\eta -\frac{\pi}{2}|\nu_S|-\frac{\pi}{4})},\\
H_{\nu2}(-\xi k\eta\gg1)\simeq \sqrt{\frac{2}{-\xi k\pi\eta }}e^{-i(-\xi k\eta -\frac{\pi}{2}|\nu_S|-\frac{\pi}{4})},
\end{eqnarray}
it is easy to find the coefficients in (\ref{approsl})
\begin{eqnarray}
a_k=\frac{\sqrt{\pi}}{2}e^{i\frac{\pi}{2}(|\nu_S|+\frac{1}{2})}~,\quad\quad\quad b_k=0,
\end{eqnarray}
such that
\begin{eqnarray}
v^S_k(\eta)=\frac{\sqrt{\pi}}{2}\sqrt{-\eta}e^{i\frac{\pi}{2}(|\nu_S|+\frac{1}{2})}H_{\nu1}(-\xi k\eta)~.\label{vssl}
\end{eqnarray}
Using another asymptotic property
\begin{eqnarray}
H_{\nu1}(-\xi k\eta\ll1)\simeq \sqrt{\frac{2}{\pi}}e^{-i\frac{\pi}{2}}2^{|\nu_S|-\frac{3}{2}}
\frac{\Gamma(|\nu|)}{\Gamma(\frac{3}{2})}(-\xi k\eta)^{-|\nu_S|} ,
\end{eqnarray}
we find
\begin{eqnarray}
v^S_k(\eta)\Big|_{-\xi k\eta\ll1}\simeq\sqrt{\frac{-\eta}{2}}(-\xi k\eta)^{-|\nu_S|} ,\label{vsslsh}
\end{eqnarray}
In order to find the possible observational effects,  it is necessary to derive the  scalar power spectrum in the super horizon limit where $-\xi k\eta\ll1$, recall the  definition of the power spectrum
\begin{eqnarray}
P_S(k)=\frac{k^3}{2\pi^2}\Big|\frac{v^S_k}{z_S}\Big|^2.\label{spsk}
\end{eqnarray}
Using the equation (\ref{Rayslr}), we have
\begin{eqnarray}
z_S=\sqrt{p}\frac{\bar{\varphi}'}{\mathcal{H}}
=\sqrt{p}\frac{\dot{\bar{\varphi}}}{H}
=\frac{\sqrt{p}}{\xi}\sqrt{\frac{2\epsilon}{\kappa}}.\label{zsexpr}
\end{eqnarray}
Substituting (\ref{vsslsh}) and  (\ref{zsexpr}) into (\ref{spsk}), we find
\begin{eqnarray}
P_S(k)\Bigg|_{-\xi k\eta\ll1}~=\frac{\kappa}{8\pi^2\xi}\frac{H^2}{\epsilon}(-\xi k\eta)^{3-2|\nu_S|}.\label{pssh}
\end{eqnarray}
From which we immediately read the amplitude of scalar perturbations
\begin{eqnarray}
A_S:=\frac{\kappa}{8\pi^2\xi}\frac{H^2}{\epsilon}
=\frac{\kappa}{8\pi^2}\frac{H^2}{\epsilon_V}\Big(1-\frac{m}{2}(3m-5)\delta_Q\Big),\label{As}
\end{eqnarray}
and the spectral index
\begin{eqnarray}
n_S-1:=3-2|\nu_S|=-6\epsilon_V\Big(1+m(m-3)\delta_Q\Big)+2\eta_V\Big(1-2m\delta_Q\Big).\label{spows}
\end{eqnarray}
\subsection{Tensor power spectrum}

In the tensor Mukhanov equation (\ref{TMukheqn}),  recall the definition of $z_T$
\begin{eqnarray}
z_T=\sqrt{\frac{p}{\mathcal{G}^{(2)}}}=\frac{\sqrt{\bar{p}}}{\xi} ,
\end{eqnarray}
using the slow roll conditions, short calculations give the coefficient
\begin{eqnarray}
\frac{z_T''}{z_T}&=&\frac{1}{\eta^2}\Big[2+3\epsilon\Big(1+\frac{1}{2} m(m^2-1)\delta_Q\Big)\Big]\nonumber\\
&=&\frac{1}{\eta^2}\Big[2+3\epsilon_V\Big(1+(\frac{1}{2} m^3+m^2-\frac{7}{2}m)\delta_Q\Big)\Big] ,
\end{eqnarray}
in the momentum space it becomes
\begin{eqnarray}
{v^T_k}''+\Big(1+m(m+1)\delta_Q\Big)k^2 v^T_k-\frac{1}{\eta^2}\Big({\nu_T}^2-\frac{1}{4}\Big)v^T_k=0 ,\label{slTMukheqn}
\end{eqnarray}
where
\begin{eqnarray}
\nu_T=\frac{3}{2}+\epsilon_V\Big(1+(\frac{1}{2} m^3+m^2-\frac{7}{2}m)\delta_Q\Big) .
\end{eqnarray}
Simply following the procedures in the last subsection, the power spectrum for tensor perturbation reads
\begin{eqnarray}
P_T(k)\Bigg|_{-\xi k\eta\ll1}:
=\frac{4\kappa k^3}{\pi^2}\Big|\frac{v^T_k}{z_T}\Big|^2\Bigg|_{-\xi k\eta\ll1}=\frac{2\kappa}{\pi^2}\frac{H^2}{\xi}(-\xi k\eta)^{3-2|\nu_T|} ,\label{psk}
\end{eqnarray}
where the amplitude and the tensor spectral index are expressed by
\begin{eqnarray}
A_T&=&\frac{2\kappa}{\pi^2}\frac{H^2}{\xi}~,\nonumber\\
n_T&:=&3-2|\nu_T|=-2\epsilon_V\Big(1+(\frac{1}{2} m^3+m^2-\frac{7}{2}m)\delta_Q\Big) .
\end{eqnarray}
The tensor-to-scalar ratio is
\begin{eqnarray}
r=\frac{A_T}{A_S}=16\epsilon_V\Big(1+\frac{m}{2}(3m-5)\delta_Q\Big) .
\end{eqnarray}

Note that in deriving the spectra indices we have ignored all the  terms  proportional to $\delta_Q^2$, $\epsilon_V^2$, $\eta_V^2$. At the same time, we treat the terms $\xi$, $\nu_S$, $\nu_T$ as constants and get the approximate solution in (\ref{approsl}). These assumptions only make sense when $\delta_Q\ll1$ but at the same time $\delta_Q>\epsilon_V$, for if $\delta_Q<\epsilon_V$ or $\eta_V$, all the above conclusions would not be of practical significance because the contributions from $\epsilon_V^2$, $\eta_V^2$ will outweigh the contributions from the terms proportional to $\epsilon\delta_Q$. In the standard $\bar{\mu}$ scheme, during the slow roll period, $\delta_Q\sim\mathcal{O}(10^{-10})$, which makes the quantum holonomy effects completely negligible, thus in order to find the potentially observable holonomy corrections, we should seek a different parametrization  of holonomy or choose a quite different initial condition.

 It is shown that in \cite{Bolliet15a,Bolliet15b} if we set the initial time in the remote past before the bounce, the vacuum state will no longer be the Bunch-Davies vacuum before the onset of inflation and the power spectra for the very short comoving scales will be drastically different from the classical theory, this result is one of the important consequences of the effective signature change. Notice again that in our article, the initial condition is set right before the onset of inflation just as in \cite{Mielczarek14a}, where $\mathcal{G}^{(2)}\sim1$ and we do not have a signature change, thus the predictions of the power spectra are quite different from those obtained in \cite{Bolliet15a,Bolliet15b}. Nevertheless, it is meaningful to briefly discuss how the signature change in this article affects the power spectra for the very short wavelengths under the initial condition used in \cite{Bolliet15a}. For simplicity, let us take the tensor perturbations for example, in the momentum space, the equation of motion (\ref{eomh}) becomes
\begin{eqnarray}
{h_k}''+2\mathcal{H}
{h_k}'-\frac{\mathcal{G}^{(3)}}{\mathcal{G}^{(2)}}(\bar{k}'-
\mathcal{H}\bar{k}){h_k}'+\mathcal{G}^{(2)}k^2h_k=0 .\label{eomhk}
\end{eqnarray}
Notice that around the bounce point, $\mathcal{G}^{(2)}$ becomes negative and $\mathcal{H}\sim0$, then for  short scales, (\ref{eomhk}) can be approximated by
\begin{eqnarray}
{h_k}''-\mid\mathcal{G}^{(2)}\mid k^2h_k\simeq0 ,
\end{eqnarray}
which leads to the approximated solution (by ignoring the variation of $\mathcal{G}^{(2)}$ around the bounce)
\begin{eqnarray}
h_k\simeq c_1\exp\left(k\int\sqrt{|\mathcal{G}^{(2)}| } d\eta\right)+c_2\exp\left(k\int-\sqrt{|\mathcal{G}^{(2)}|} d\eta\right).
\end{eqnarray}
Since $P_T(k)\propto|h_k|^2$, it is obvious that the power spectrum will also receive an exponential growth near the bounce. If the bounce phase lasts long enough or $|\mathcal{G}^{(2)}|$ has large value near the bounce, it is probable that the integral will grow rapidly and spoils the near scale invariant classical predictions for short comoving wavelengths. The concrete predictions of the power spectra by using an explicit formulation of the holonomy $g(\bar{k},\bar{p})$ (such as the ones proposed in \cite{Achour17}) under this initial condition are worth investigating in detail in future work.
\section{Summary and Remarks}

In this paper, motivated by the possibility that the holonomy correction could take different form from the standard $\bar{\mu}$ scheme in LQC, we replace the classical connection variable $\bar{k}$ by a general, undetermined function $g(\bar{k},\bar{p})$ in the effective Hamiltonian, then by requiring an anomaly free algebra, we find an interesting result, which says $g(\bar{k},\bar{p})$ could only take the form $\bar{p}^{\frac{1}{2}}f(\frac{\bar{k}}{\bar{p}^{\frac{1}{2}}})$ where $f$ is an arbitrary function which allows $g(\bar{k},\bar{p})=\bar{k}$ in the classical limit. This result agrees with the previous results obtained in \cite{Cailleteau12a} for standard $\bar{\mu}$ scheme. After fixing the counter terms, we obtain the equations of motion for gauge invariant scalar, tensor and vector perturbations as well as the Mukhanov equations. Moreover,  we apply these results to study the slow roll inflation, by using some approximations and the same initial conditions as in the classical theory, we derive the scalar and tensor spectral indices and the tensor-to-scalar ratio with quantum holonomy corrections. Before we end this paper, let us make a few more remarks,

1. In this paper, when calculating the constraint algebras, Gauss constraint has also been included, which can serve the purpose of cross checking the results but is usually missed in previous literatures; Moreover, in previous works, the term $Z^{cidj}_{ab}~( \partial_c \delta E^a_i) (\partial_d \delta E^b_j) $ is usually decomposed into different expressions for scalar, tensor and vector modes before calculating the algebra and then each of these modes is shown to satisfy the same Poisson bracket. In this paper, due to the commuting relation (\ref{bracketZk}), the Poisson algebra (\ref{HHbrac})  naturally hold for all kinds of perturbations without doing the decomposition first.
In addition, we have found that in the spatially flat case all counter terms can be uniquely fixed except for  $\alpha_3$ and $\alpha_9$, It is worth mentioning that this important fact was first pointed out in \cite{Cailleteau14} for standard $\bar{\mu}$ scheme, actually, if these two terms are non-zero, many important equations and conclusions will be changed, for instance,  the signature change in the $\bar{\mu}$ scheme will not happen if we allow $1+\alpha_3<0$ near the bounce. In this paper, as usually done in previous works, we artificially assume these two terms to be zero, and in our companion paper we will prove this assumption from a different perspective.

2. It is easy to check the gauge invariant equations for scalar and tensor perturbations agree with the results obtained in \cite{Cailleteau12a,Cailleteau12c} for standard $\bar{\mu}$ scheme, however, the equation (\ref{eomSigma}) for gauge invariant vector perturbation does not agree with the previous results obtained in \cite{Mielczarek12a} because in this paper we do not put the restrictions such as $\delta_a^i\delta E^a_i=0$ for vector modes in the constraints before calculating the constraint algebra, and it is only after we solve the counter terms that we  separately put different restrictions on the corresponding modes in the constraints and then derive their equations of motion, which is opposite of what has been done in \cite{Mielczarek12a}. From these perturbation equations, one easily see that around the point $\mathcal{G}^{(2)}=0$ they are not well behaved and some variables are ill-defined, this fact means we need some delicate skills to deal with the problems around this point.

3. Finally, cautions should be taken because we only use the point-wise holonomy in this paper,  concretely speaking, as pointed out in \cite{Bojowald14}, when considering the holonomy corrections, terms representing the derivatives of the connection like $\partial K_a^i$ should also be considered in the perturbed constraints. Thus in this paper we are actually using a minimal approach by artificially setting the coefficients before these terms to zero.  Needless to say, obtaining a consistent algebra after including these terms is tough work and worth investigating in the future. On the other hand, it is worth remarking that the results obtained in this paper is not only limited to effective LQC, becasue it is possible that in some other quantum cosmology models the extrinsic curvature could also take quantum corrections like in this paper on the effective level,  if we assume these corrections also satisfy an anomaly free constraint algebra, then the arguments in this paper are still valid. We hope that in some of these models the quantum correction parameter $\delta_Q$ could be much larger than that in LQC such that the results in section $4$ could in principle be detected in the not very far future.

\ack
 This work is supported by the National Natural Science Foundation of China (NSFC) under grants No.11647068, No.11475143 and Nanhu Scholars Program for Young Scholars of Xinyang Normal University.

\end{document}